\documentclass{iopart}

\usepackage{iopams}
\usepackage{amssymb}
\usepackage{graphicx}
\usepackage{upgreek}
\usepackage{bm}

\newcommand{\sprod}{\!\cdot\!}
\newcommand{\vprod}{\!\times\!}
\newcommand{\tprod}{\!\otimes\!}
\newcommand{\conv}{\!\star\!}

\begin{document}
\title{Casimir-Polder Forces between Chiral Objects}

\author{David T. Butcher$^1$, Stefan Yoshi Buhmann$^1$ and Stefan
Scheel$^{1,2}$} 
\address{$^1$ Quantum Optics and Laser Science, Blackett Laboratory,
Imperial College London, Prince Consort Road, London SW7 2AZ, United
Kingdom}
\address{$^2$ Institut f\"ur Physik, Universit\"at Rostock,
Universit\"atsplatz 3, D-18051 Rostock, Germany}
\ead{david.butcher05@imperial.ac.uk}

\date{\today}

\begin{abstract}
The chiral component of the Casimir-Polder potential is derived within
the framework of macroscopic quantum electrodynamics. It is shown to
exist only if the particle and the medium are both chiral.
Furthermore, the chiral component of the Casimir-Polder potential can
be attractive or repulsive, depending on the chirality of the molecule
and the medium. The theory is applied to a cavity geometry in the
non-retarded limit with the intention of enantiomer separation. For a
ground state molecule the chiral component is dominated by the
electric component and thus no explicit separation will happen. If the
molecule is initially in an excited state the electric component of
the Casimir-Polder force can be suppressed by an appropriate choice of
material and the chiral component can select the molecule based on its
chirality, allowing enantiomeric separation to occur. 
\end{abstract}


\maketitle

\tableofcontents

\section{Introduction}

A three-dimensional object that cannot be superimposed on its mirror
image is said to be chiral, these distinct mirror images are called
enantiomers. Spectroscopically, enantiomers have identical properties
and distinguishing between the two is not trivial. The characteristic
feature of chiral objects is the manner of their interactions with
other chiral objects. For example, the refractive indices of left- and
right-handed circularly polarised light are different in a chiral
medium, therefore the two polarisations will propagate at different
speeds. The difference in velocity is related to the phenomenon of
circular dichroism, where the wave with the `slower' polarisation is
absorbed more strongly as it travels through the medium
\cite{condon37}.

Many of the processes crucial to life involve chiral molecules whose
chiral identity plays a central role in their chemical reactions, the
incorrect enantiomer reacting differently and not producing the
required result. In nature these molecules only occur as one 
enantiomer and are not found as the other, thus the reactions only 
occur when the correct enantiomer is present. In contrast, artificial 
production creates both enantiomers in equal proportions. Therefore it 
is important to be able to distinguish between enantiomers and 
ultimately to be able to separate a racemic (containing both enantiomers) 
mixture into an enantiomerically pure sample.

A frequently used method to separate enantiomers in an industrial
setting is chiral chromotography. The initial racemic solution is
passed through a column packed with a resolving agent, which is
usually an enzyme and by necessity has to be chiral. This either
retards, or stops, the progress of one of the enantiomers passing
through the column but crucially not for the other and thus allows the
solution to be separated. From an optical viewpoint it has been
proposed that a racemic sample can be purified by use of coordinated
laser pulses, which use a two-step process to initially drive
different transitions in the enantiomers before converting one into
the other \cite{kral_two_step}. It has recently been calculated that
in the presence of a chiral carbon nanotube the enantiomers of alanine
possess different absorption energies and it was theorised that this
could lead to a method of discrimination \cite{nanotubes}.
Furthermore, it has been shown that the van der Waals dispersion force
between molecules can be enantiomer selective
\cite{hund_09,craig_mqed}. Here we propose the use of another
dispersion force, the Casimir-Polder force \cite{casimir_polder}, as a
method of distinguishing and ultimately separating enantiomers.

Casimir-Polder forces occur due to fluctuations in the quantum vacuum
between objects, in this case it is between a particle and a
macroscopic body. This force can be decomposed into an electric
component that is solely dependent upon the electric dipole moments of
the particle, the magnetic component of the force is likewise defined.
We show that there is a third component that depends upon an interaction 
of the molecule's electric and magnetic dipole moments and this 
describes the particles chiral response in the chiral component of the 
Casimir-Polder force. For the purpose of discrimination, the force on 
enantiomers must be different, this means that a chiral medium is required 
to produce the chiral-selective behaviour.

Chiral molecules occur frequently in nature and there has been much
work to calculate the relevant transition values either ab initio
\cite{rauk84,crawford06}, or through a twisted arc model for simple
chiral molecules \cite{trost_twistedarc}. Recent technological
developments have allowed for the creation of chiral fullerenes such
as C$_{76}$ \cite{c76_whetten}, chiral carbon nanotubes
\cite{nanotubes} and chiral meta-materials \cite{meta_review}. Chiral
meta-materials can be made from structures such as a gold helix
\cite{chiral_helix}, which shows a broadband electromagnetic response;
a woodpile structure \cite{rillopplett09}, a gold bar construction
\cite{zhang_09}, which exhibit a negative refractive index in certain
frequency ranges; and gold dots, \cite{chiral_oligomers} which can be
tuned during construction to give a desired response. Recently it has
been shown that superchiral electromagnetic fields
\cite{optical_chirality} can arise in planar chiral metamaterials.
These fields can generate a much larger dissymmetry between the
effective refractive indices for adsorbed chiral molecules on left-
and right-handed materials than circularly polarised light and a
solution of chiral molecules \cite{superchiral_fields}. Chiral
metamaterials have also been proposed as a method of producing
repulsive Casimir forces \cite{zhao09}.

In this article we derive the chiral component of the Casimir-Polder
potential within the framework of macroscopic quantum electrodynamics
\cite{scheelbuhmann08}. We show that inherent to this potential is the
requirement that the particle and the medium both exhibit a chiral
response. The theory is applied to the case of a chiral molecule in a
cavity between two chiral metamaterials in the non-retarded limit,
where we show a possible route towards enantiomer separation.

The article is organised as follows. The quantised description of a
chiral medium is given in section \ref{secchiral}, followed by the
derivation of the chiral component of the Casimir-Polder force in
section \ref{seccpchiral}. This is applied to the case of a chiral
molecule near a boundary; a perfect chiral mirror is considered in
section \ref{secperfect} and a chiral metamaterial in section
\ref{secmeta}. We illustrate the theory by examining a chiral molecule
in a cavity made of chiral metamaterials, considering a molecule
initially in the ground state in section \ref{secground} and initially
in an excited state in section \ref{secexcited}. We provide some
concluding remarks in section \ref{secconclusion}.


\section{Formalism}
In order to derive the chiral component of the Casimir-Polder
potential in a general form, field quantisation in an absorbing chiral
medium is required. This quantisation is the basis for describing the
electric and magnetic fields and the subsequent formulation of an
important integral relation, which is needed for the calculation of the
Casimir-Polder potential.

\subsection{Chiral Media}
\label{secchiral}
The constitutive  relations for the electromagnetic fields in the
presence of a classical, non-dissipative chiral medium are given by
\begin{equation}\label{eqconrelD}
\mathbf{D} = \varepsilon_0\bm{\varepsilon}\conv\mathbf{E}
- \frac{i}{c}\bm{\kappa}^{\mathrm{T}}\conv\mathbf{H} 
\end{equation}
and
\begin{equation}\label{eqconrelB}
\mathbf{B} =  \frac{i}{c}\bm{\kappa}\conv\mathbf{E}
+ \mu_0\bm{\mu}\conv\mathbf{H}
\end{equation}
where $\bm{\varepsilon}$ and $\bm{\mu}$ are the relative permittivity
and permeability tensors and $\bm{\kappa}$ is the chirality tensor, a
magnetoelectric susceptibility. The chiral susceptibility of the
medium has the effect of `rotating' a magnetic effect to contribute
towards an electric response and vice versa. The $\conv$ is a
notational shorthand for a spatial convolution, i.e.,
$[\bm{X}\conv\bm{Y}](\mathbf{r}, \mathbf{r}') = \int
\mathrm{d}^{3}\mathbf{s}\ \bm{X}(\mathbf{r}, \mathbf{s})\sprod
\bm{Y}(\mathbf{s}, \mathbf{r}')$. For nonlocally responding media the
permittivity, permeability and chiral susceptibilities take the form
$\bm{\varepsilon}(\mathbf{r}, \mathbf{r}', \omega)$, etc., where the
two spatial variables are independent. This reduces to the form
$\bm{\varepsilon}(\mathbf{r},
\omega)\delta(\mathbf{r}\!-\!\mathbf{r}')$ in a locally responding
medium. The medium is assumed to be reciprocal,
i.e. $\bm{\varepsilon}(\mathbf{r}, \mathbf{r}') =
\bm{\varepsilon}^{\mathrm{T}}(\mathbf{r}',\mathbf{r})$ and
$\bm{\mu}(\mathbf{r}, \mathbf{r}') =
\bm{\mu}^{\mathrm{T}}(\mathbf{r}',\mathbf{r})$.

The Casimir-Polder force requires field quantisation in a medium due
to its explicit quantum nature. In an absorbing medium, the quantum
version of the constitutive relations are given as \cite{njp_2012}
\begin{equation}\label{eqconstitD}
\hat{\mathbf{D}}
=\varepsilon_0\bm{\varepsilon}\conv\hat{\mathbf{E}}
- \frac{i}{c}\bm{\kappa}^{\mathrm{T}}\conv\hat{\mathbf{H}} +
\hat{\mathbf{P}}_\mathrm{N}
-\frac{i}{c}\bm{\kappa}^{\mathrm{T}}\conv\hat{\mathbf{M}}_\mathrm{N}
\end{equation} and
\begin{equation}\label{eqconstitB}
\hat{\mathbf{B}} =  \frac{i}{c}\bm{\kappa}\conv\hat{\mathbf{E}}
+ \mu_0\bm{\mu}\conv\hat{\mathbf{H}} +
\mu_0\bm{\mu}\conv\hat{\mathbf{M}}_\mathrm{N}.
\end{equation} 
The terms $\hat{\mathbf{P}}_\mathrm{N}$ and
$\hat{\mathbf{M}}_\mathrm{N}$ are the noise polarisation and
magnetisation respectively, they describe the dissipation in the
medium and form a Langevin noise current
\begin{equation}\label{eqnoisej}
\hat{\mathbf{j}}_\mathrm{N}(\mathbf{r}, \omega)
= -i\omega\hat{\mathbf{P}}_\mathrm{N}(\mathbf{r}, \omega) +
\mathbf{\nabla}\vprod\hat{\mathbf{M}}_\mathrm{N}(\mathbf{r}, \omega).
\end{equation}
The wave equation for the electric field in the chiral medium is then
\begin{eqnarray}
\Bigl[\mathbf{\nabla}\vprod\bm{\mu}^{-1}\conv\mathbf{\nabla}\vprod
+ \frac{\omega}{c}(\mathbf{\nabla}\vprod\bm{\mu}^{-1}\conv\bm{\kappa}
+ \bm{\kappa}^{\mathrm{T}}\conv\bm{\mu}^{-1}
\conv\mathbf{\nabla}\vprod)
\nonumber \\ -\frac{\omega^2}{c^2}(\bm{\varepsilon} -
\bm{\kappa}^{\mathrm{T}}\conv\bm{\mu}^{-1}\conv\bm{\kappa})\Bigl]\,
\conv\,\hat{\mathbf{E}} = i\omega\mu_{0}\hat{\mathbf{j}}_\mathrm{N}
\end{eqnarray} 
and the Green's tensor, $\bm{G}(\mathbf{r}, \mathbf{r}', \omega)$, is 
the fundamental solution to the above inhomogeneous Helmholtz
equation 
\begin{eqnarray}
\Bigl\{\Bigl[\mathbf{\nabla}\vprod\bm{\mu}^{-1}
\conv\mathbf{\nabla}\vprod +
\frac{\omega}{c}(\mathbf{\nabla}\vprod\bm{\mu}^{-1}\conv\bm{\kappa} +
\bm{\kappa}^{\mathrm{T}}\conv\bm{\mu}^{-1}\conv\mathbf{\nabla}\vprod )
\nonumber \\-\frac{\omega^2}{c^2}(\bm{\varepsilon} -
\bm{\kappa}^{\mathrm{T}}\conv\bm{\mu}^{-1}\conv\bm{\kappa})\Bigl]\,
\conv\,\bm{G}\Bigl\}(\mathbf{r}, \mathbf{r}') =
\bm{\delta}(\mathbf{r}\!-\!\mathbf{r}').
\end{eqnarray} 
The Green's tensor obeys the Schwarz reflection principle,
\begin{equation}\label{eqschwarz}
\bm{G}^{*}(\mathbf{r}, \mathbf{r}', \omega)
 = \bm{G}(\mathbf{r}, \mathbf{r}', -\omega^{*}),
\end{equation}
which ensures reality of $\bm{G}(\mathbf{r}, \mathbf{r}', t)$, and
inherits compliance with the Onsager condition for reciprocal media
from the medium response functions
\begin{equation}\label{eqonsager}
\bm{G}(\mathbf{r}, \mathbf{r}', \omega)
= \bm{G}^{\mathrm{T}}(\mathbf{r}', \mathbf{r}, \omega).
\end{equation}
The wave equation and its Green's-tensor solution uniquely define
the electric field (in coordinate space) as
\begin{equation}\label{eqelectricfield}
\hat{\mathbf{E}}(\mathbf{r}, \omega)
= i\omega\mu_{0}\int \mathrm{d}^{3}\mathbf{r}' \bm{G}(\mathbf{r},
\mathbf{r}', \omega)\sprod \hat{\mathbf{j}}_\mathrm{N}(\mathbf{r}',
\omega)
\end{equation}
and the magnetic induction field as
\begin{equation}\label{eqmagneticfield}
\hat{\mathbf{B}}(\mathbf{r}, \omega)
= \mu_{0}\int\mathrm{d}^{3}\mathbf{r}'
\mathbf{\nabla}\vprod\bm{G}(\mathbf{r}, \mathbf{r}',
\omega)\sprod\hat{\mathbf{j}}_\mathrm{N}(\mathbf{r}', \omega). 
\end{equation}

To describe a general dissipative chiral medium, it is necessary to 
introduce the fundamental degrees of freedom of the field--medium system.
Starting with the commutation relations for the
noise polarisation and magnetisation \cite{njp_2012}
\begin{eqnarray}\label{eqjnjn}
&&[\hat{\mathbf{P}}_{\mathrm{N}}(\mathbf{r}, \omega),
\hat{\mathbf{P}}^{\dag}_{\mathrm{N}}(\mathbf{r}', \omega')] \nonumber \\ &&=
\frac{\varepsilon_{0}\hbar}{\pi} \bigl\{\mathrm{Im}[\bm{\varepsilon}(\omega)\!-\!
\bm{\kappa}^{\mathrm{T}}(\omega)\conv \bm{\mu}^{-1}(\omega)\conv
\bm{\kappa}(\omega)]\bigr\}
(\mathbf{r},\mathbf{r}')
\delta(\omega\!-\!\omega'), 
\end{eqnarray}\begin{eqnarray}\label{eqjnmn}
&&[\hat{\mathbf{P}}_{\mathrm{N}}(\mathbf{r}, \omega),
\hat{\mathbf{M}}^{\dag}_{\mathrm{N}}(\mathbf{r}', \omega')] \nonumber \\ &&=
\frac{\hbar}{iZ_{0}\pi} \bigl\{\mathrm{Im}[\bm{\kappa}^{\mathrm{T}}(\omega)\conv
\bm{\mu}^{-1}(\omega)] \bigr\}
(\mathbf{r},\mathbf{r}')
\delta(\omega\!-\!\omega'),
\end{eqnarray}\begin{eqnarray}\label{eqmnjn}
&&[\hat{\mathbf{M}}_{\mathrm{N}}(\mathbf{r}, \omega),
\hat{\mathbf{P}}^{\dag}_{\mathrm{N}}(\mathbf{r}', \omega')] \nonumber \\ &&= -
\frac{\hbar}{iZ_{0}\pi} \bigl\{\mathrm{Im}[\bm{\mu}^{-1}(\omega)\conv
\bm{\kappa}(\omega)]\bigr\}
(\mathbf{r},\mathbf{r}')
\delta(\omega\!-\!\omega'), 
\end{eqnarray}\begin{equation}\label{eqmnmn}
[\hat{\mathbf{M}}_{\mathrm{N}}(\mathbf{r}, \omega),
\hat{\mathbf{M}}^{\dag}_{\mathrm{N}}(\mathbf{r}', \omega')] =
-\frac{\hbar}{\mu_{0}\pi}
\mathrm{Im}[\bm{\mu}^{-1}
(\mathbf{r},\mathbf{r}',\omega)]
\delta(\omega\!-\!\omega'), 
\end{equation}
the noise polarisation and magnetisation can be decomposed into 
\begin{equation}\label{eqnoisecurr}
\left(\begin{array}{cc}\hat{\mathbf{P}}_\mathrm{N}\\
\hat{\mathbf{M}}_\mathrm{N}\end{array}\right)
 =\sqrt{\frac{\hbar}{\pi}}\,\mathcal{R}\conv
\left( \begin{array}{cc}\hat{\mathbf{f}}_{e}\\
 \hat{\mathbf{f}}_{m}\end{array}\right)
\end{equation}
where $\mathcal{R}$ is the `square root'
of the $6 \mathrm{x} 6$ response tensor,
\begin{equation}\label{eqrrdagmatrix}
\mathcal{R} \conv \mathcal{R}^{\dag}
= \left( \begin{array}{cc}\varepsilon_{0}
\mathrm{Im}[\bm{\varepsilon}\!-\! \bm{\kappa}^{\mathrm{T}}\conv
\bm{\mu}^{-1}\conv \bm{\kappa}] &
\frac{\mathrm{Im}[\bm{\kappa}^{\mathrm{T}}\conv\,
\bm{\mu}^{-1}]}{iZ_{0}} \\ - \frac{\mathrm{Im}[\bm{\mu}^{-1}\conv\,
\bm{\kappa}]}{iZ_{0}} & - \frac{\mathrm{Im}[\bm{\mu}^{-1}]}{\mu_{0}}
\end{array} \right),
\end{equation}
which describes the dissipative properties of the medium, where $Z_{0} 
= \sqrt{\frac{\mu_{0}}{\varepsilon_{0}}}$. For a
passive, isotropic medium the response functions are restricted by
$(\mathrm{Im}[\kappa])^{2} < \mathrm{Im}[\varepsilon]\mathrm{Im}[\mu]$
\cite{guerin_94}. The vector fields $\hat{\mathbf{f}}_{e}(\mathbf{r},
\omega)$ and $\hat{\mathbf{f}}_{m}(\mathbf{r}, \omega)$ are the
bosonic annihilation operators for the matter-electromagnetic field
system and with the creation operators they obey the commutation
relation
\begin{equation}
[\hat{\mathbf{f}}_{\lambda}(\mathbf{r}, \omega),
\hat{\mathbf{f}}^{\dag}_{\lambda'}(\mathbf{r}', \omega')] =
\delta_{\lambda\lambda'}\bm{\delta}(\mathbf{r}\!-\!\mathbf{r}
')\delta(\omega\!-\!\omega') 
\end{equation}with $\lambda, \lambda' = e,m$.
By rewriting the electric and magnetic fields as
\begin{equation}\label{eqlambdaefield}
\hat{\mathbf{E}}(\mathbf{r},\omega)\!=\!\sum_{\lambda = e,m}\!
\int\mathrm{d}^{3}\mathbf{r}' \bm{G}_{\lambda}(\mathbf{r},
\mathbf{r}', \omega)\sprod \hat{\mathbf{f}}_{\lambda}(\mathbf{r}',
\omega),
\end{equation}
\begin{equation}\label{eqlambdabfield}
\hat{\mathbf{B}}(\mathbf{r}, \omega)\! =\! \sum_{\lambda = e,m}\!
\int\mathrm{d}^{3}\mathbf{r}'\mathbf{\nabla}\vprod
\bm{G}_{\lambda}(\mathbf{r}, \mathbf{r}', \omega)\sprod
\hat{\mathbf{f}}_{\lambda}(\mathbf{r}', \omega) 
\end{equation} 
and making use of (\ref{eqnoisecurr}) and (\ref{eqrrdagmatrix}) it can
be shown that
\begin{equation}\label{eqgegm}
\left(\begin{array}{c}\bm{G}_{e}(\mathbf{r}, \mathbf{r}',\omega) \\ \bm{G}_{m}
(\mathbf{r}, \mathbf{r}',\omega) \end{array}\right)
= -i\mu_{0}\omega\sqrt{\frac{\hbar}{\pi}} [\bm{G}(\omega) \conv
\left(\begin{array}{c}i\omega \\
\vprod\overleftarrow{\mathbf{\nabla}}\end{array}\right) \sprod
\mathcal{R}(\omega)](\mathbf{r}, \mathbf{r}')
\end{equation}
where the operation $\vprod\overleftarrow{\mathbf{\nabla}}$ refers to
taking the derivatives of the second spatial variable and
mathematically is described as $[\bm{T}\vprod
\overleftarrow{\mathbf{\nabla}}]_{ij}(\mathbf{r},\mathbf{r}')\!=\!
\epsilon_{jkl}\partial_l'T_{ik}(\mathbf{r},\mathbf{r}')$. 

We can now derive the following integral relation,
\begin{equation}\label{intG}
\sum_{\lambda = e,m}\int\!\mathrm{d}^{3}\mathbf{s}
\bm{G}_{\lambda}(\mathbf{r}, \mathbf{s},
\omega)\sprod\bm{G}^{\dag}_{\lambda}(\mathbf{r}', \mathbf{s}, \omega)
\!=\!\frac{\hbar\mu_{0}\omega^{2}}{\pi} \mathrm{Im}\bm{G}(\mathbf{r},
\mathbf{r}', \omega),
\end{equation}
see \ref{appintrel}.


\subsection{Chiral Casimir-Polder Potential}
\label{seccpchiral}
To derive the Casimir-Polder potential we start with the interaction
Hamiltonian in multipolar coupling and long wavelength approximation,
describing the interaction of a molecule with the electric and
magnetic fields  \cite{scheelbuhmann08}
\begin{equation}\label{eqcpham}
\hat{H}_{AF}= -\hat{\mathbf{d}}\sprod\hat{\mathbf{E}}(\mathbf{r}_{A})
- \hat{\mathbf{m}}\sprod \hat{\mathbf{B}}(\mathbf{r}_{A}),
\end{equation}
where diamagnetic interactions have been neglected. Initially the
particle can be either in its ground state or an excited state and the
energy shift due to the atom-field interaction, (\ref{eqcpham}), is
given by
\begin{equation}\label{eqgroundpert}
\Delta E = \sum_{I\neq N} \frac{\langle N
\vert\hat{H}_{AF}\vert I\rangle\langle I \vert \hat{H}_{AF}\vert
N\rangle}{E_{N} - E_{I}}.
\end{equation}
When multiplying out the matrix elements it is important to note that
the cross terms involving both electric and magnetic dipole
interactions do not vanish and are in fact responsible for the chiral
interaction, i.e., $\langle N \vert -
\hat{\mathbf{d}}\sprod\hat{\mathbf{E}}(\mathbf{r}_{A})\vert
I\rangle\langle I \vert -\hat{\mathbf{m}}\sprod
\hat{\mathbf{B}}(\mathbf{r}_{A})\vert N\rangle$ and $\langle N \vert
-\hat{\mathbf{m}}\sprod\hat{\mathbf{B}}(\mathbf{r}_{A})\vert
I\rangle\langle I \vert -\hat{\mathbf{d}}\sprod
\hat{\mathbf{E}}(\mathbf{r}_{A}) \vert N\rangle\neq 0$.

The initial state is denoted by $\vert N\rangle = \vert
n\rangle\vert\{0\}\rangle$ and the intermediate states $\vert I\rangle
= \vert k\rangle\hat{\mathbf{f}}^{\dagger}_{\lambda}(\mathbf{r},
\omega)\vert\{0\}\rangle$ where $\vert\{0\}\rangle$ denotes the ground
state of the matter-field system (upon which the bosonic creation
operators $\hat{\mathbf{f}}^{\dagger}_{\lambda} (\mathbf{r}, \omega)$ 
act) and $\vert n\rangle$ and
$\vert k\rangle$ represent the initial and intermediate energy levels
of the particle. The summation in (\ref{eqgroundpert}) contains sums
over molecular transitions, the polarisation modes of the
electromagnetic fields and integrals over all space and positive
frequencies
\begin{equation*}
\sum_{I \neq N} \rightarrow
\sum_{k}\sum_{\lambda = e,m}\int\mathrm{d}^{3}\mathbf{r}
\,\mathcal{P}
\int^{\infty}_{0}
\mathrm{d}\omega
\end{equation*}
($\mathcal{P}$: principal value).
Using the definitions of the electric (\ref{eqlambdaefield}) and
magnetic induction (\ref{eqlambdabfield}) fields it can be shown that
\begin{equation} 
\langle N \vert - \hat{\mathbf{d}}\sprod\hat{\mathbf{E}}
(\mathbf{r}_{A})\vert I\rangle = - \mathbf{d}_{nk}\sprod
\bm{G}_{\lambda} (\mathbf{r}_{A}, \mathbf{r}, \omega),
\end{equation}
\begin{equation} 
\langle N \vert - \hat{\mathbf{m}}\sprod\hat{\mathbf{B}}
(\mathbf{r}_{A})\vert I\rangle = -
\frac{\mathbf{m}_{nk}\sprod\mathbf{\nabla}\vprod\bm{G}_{\lambda}
(\mathbf{r}_{A}, \mathbf{r}, \omega)}{i\omega},
\end{equation} 
where the electric dipole transition matrix elements are $\langle
n\vert\hat{\mathbf{d}}\vert k\rangle = \mathbf{d}_{nk}$ and $\langle
n\vert\hat{\mathbf{m}}\vert k\rangle = \mathbf{m}_{nk}$ the
corresponding magnetic dipole moment matrix elements.

The electric and magnetic components of the Casimir-Polder potential
are well known and the result can be found in \cite{scheelbuhmann08}.
Here we focus on the terms containing a single curl operation and a
dependency on the electric and magnetic transition matrix elements as
it is these terms that give rise to the chiral component of the
Casimir-Polder potential:
\begin{eqnarray}
\Delta E_{c} = \frac{\mu_0}{\pi}\sum_k \mathcal{P} \int^{\infty}_{0}
\frac{i\ \mathrm{d}\omega\ \omega}{\omega_{kn}\!+\!\omega}
\Bigl[\mathbf{d}_{nk}\sprod\mathrm{Im}[\bm{G}(\mathbf{r}_{A},
\mathbf{r}_{A},
\omega)]\vprod\overleftarrow{\mathbf{\nabla}}'\sprod\mathbf{m}_{kn}
\nonumber \\ +
\mathbf{m}_{nk}\sprod\mathbf{\nabla}\vprod\mathrm{Im}[\bm{G}(\mathbf{r
}_{A}, \mathbf{r}_{A}, \omega)]\sprod\mathbf{d}_{kn} \Bigl],
\end{eqnarray}
the transition frequencies are defined as $\omega_{kn} = \omega_{k}
\!-\! \omega_{n}$. The Casimir-Polder potential is the position
dependent part of the total energy shift,
$\Delta E = \Delta E_0+U(\mathbf{r}_{A})$,
and so its only contribution arises
from the scattering part of the Green's function,
$\bm{G}^{(1)}(\mathbf{r}, \mathbf{r}', \omega)$.

The imaginary part of the Green's tensor is written as
$\mathrm{Im}[\bm{G}(\mathbf{r}_{A}, \mathbf{r}_{A}, \omega)] =
\frac{1}{2i}[\bm{G}(\mathbf{r}_{A}, \mathbf{r}_{A}, \omega)-
\bm{G}^{*}(\mathbf{r}_{A}, \mathbf{r}_{A}, \omega)]$. When the
molecule is not initially in the ground state care needs to be taken
of the poles that occur for transitions to states of lower energy than
the initial state. By using contour integration techniques the
off-resonant part of the Casimir-Polder potential can be obtained in
terms of an integral over imaginary frequencies, whereas the resonant
part is due to the residue at the poles. The result is

\begin{eqnarray}\label{eqfullchiralcp}
\Delta E_{c}\! =\!
-\frac{\mu_0\hbar}{2\pi}\int^{\infty}_{0}\mathrm{d}\xi\
\xi\bigl(\mathrm{tr}[\Gamma_{em}(i\xi)\sprod\bm{G}^{(1)}(\mathbf{r}_{A
}, \mathbf{r}_{A}, i\xi) \vprod \overleftarrow{\mathbf{\nabla}}']
\nonumber \\ + \mathrm{tr}[\Gamma_{me}(i\xi)\sprod \mathbf{\nabla}
\vprod \bm{G}^{(1)}(\mathbf{r}_{A}, \mathbf{r}_{A}, i\xi)] \bigl)
\nonumber \\ + i\mu_{0} \sum_{k} \Theta(\omega_{nk})
\omega_{nk}\bigl(\mathbf{d}_{nk} \sprod
\mathrm{Re}[\bm{G}^{(1)}(\mathbf{r}_{A}, \mathbf{r}_{A}, \omega_{nk})]
\vprod \mathbf{\nabla}' \sprod \mathbf{m}_{kn} \nonumber \\ +
\mathbf{m}_{nk} \sprod \mathbf{\nabla}\vprod
\mathrm{Re}[\bm{G}^{(1)}(\mathbf{r}_{A}, \mathbf{r}_{A}, \omega_{nk})]
\sprod \mathbf{d}_{kn} \bigl), 
\end{eqnarray}
where 
\begin{equation}\label{eqgammaem}
\Gamma_{em} (i\xi) =
\frac{1}{\hbar}\sum_k\left(\frac{\mathbf{m}_{kn}\otimes\mathbf{d}_{nk}
}{\omega_{kn}\!+\!i\xi} -
\frac{\mathbf{m}_{kn}\tprod\mathbf{d}_{nk}}{\omega_{kn}\!-\!i\xi}
\right),
\end{equation}\begin{equation}\label{eqgammame}
\Gamma_{me} (i\xi) =
\frac{1}{\hbar}\sum_k\left(\frac{\mathbf{d}_{kn}\otimes\mathbf{m}_{nk}
}{\omega_{kn}\!+\!i\xi} -
\frac{\mathbf{d}_{kn}\tprod\mathbf{m}_{nk}}{\omega_{kn}\!-\!i\xi}
\right)
\end{equation}
are chiral susceptibility tensors and $\Theta(\omega_{nk})$ is a step
function which is zero if $\omega_{nk} < 0$, i.e. for a transition to
a higher energy state. For an isotropic particle
($\mathbf{d}\tprod\mathbf{m} = \frac{\mathbf{d}\sprod\mathbf{m}}{3}$),
(\ref{eqgammaem}) and (\ref{eqgammame}) both reduce to
\begin{equation}
\Gamma(i\xi) =  - \frac{2}{3\hbar}\sum_{k} \frac{\xi\
R_{nk}}{(\omega_{kn})^2\!+\!\xi^2}.
\end{equation}
The term $R_{nk}$ is the optical rotatory strength and is defined as 
\begin{equation}
R_{nk} = \mathrm{Im}(\mathbf{d}_{nk}\sprod\mathbf{m}_{kn}).
\end{equation} 
It describes the interaction between electric and magnetic dipole
moments in a molecule and, crucially, left and right -handed
enantiomers differ in the sign of $R_{nk}$ for a particular
transition.

To further simplify (\ref{eqfullchiralcp}), it can be shown that
\begin{equation}\label{eqtrg}
\mathrm{tr}[\bm{G}(\mathbf{r}_{A},
\mathbf{r}_{A}, \omega)\vprod\overleftarrow{\mathbf{\nabla}}'] = -
\mathrm{tr}[\mathbf{\nabla}\vprod\bm{G}(\mathbf{r}_{A},
\mathbf{r}_{A}, \omega)]
\end{equation}
where the Onsager reciprocity relation~(\ref{eqonsager}) has been
used. For the real part of the Green's function an equivalent
expression holds. We can now write the chiral component of the
Casimir-Polder potential as
\begin{eqnarray}\label{eqchiralcp}
U_{c}(\mathbf{r}_{A})
= - \frac{\hbar\mu_0}{\pi} \int^{\infty}_{0}\mathrm{d}\xi\ \xi\
\Gamma(i\xi) \mathrm{tr}[\mathbf{\nabla}\vprod\bm{G}^{(1)}
(\mathbf{r}_{A}, \mathbf{r}_{A}, i\xi)] \nonumber
\\+\frac{2\mu_{0}}{3} \sum_{k}\Theta(\omega_{nk})\omega_{nk} R_{nk}
\mathrm{tr}[\mathbf{\nabla} \vprod
\mathrm{Re}[\bm{G}^{(1)}(\mathbf{r}_{A}, \mathbf{r}_{A},
\omega_{nk})]]
\end{eqnarray} 
The requirement that a chiral object is needed to be able to
distinguish between the chiral states of another object is fulfilled.
As (\ref{eqchiralcp}) shows we must have a chiral medium and a chiral
molecule. If either the medium
($\mathrm{tr}[\mathbf{\nabla}\vprod\bm{G}^{(1)}(\mathbf{r}_{A},
\mathbf{r}_{A}, i\xi)] = 0$), or the particle ($R_{nk} = 0$) is
achiral there will not be a chiral component to the Casimir-Polder
potential. This can be thought of as a generalisation to the Curie
dissymmetry principle (originally formulated for crystal symmetries)
and one may say: The Casimir-Polder potential cannot distinguish
between molecules of different handedness if the medium does not
possess chiral properties itself.

For reference, the off-resonant electric and magnetic Casimir-Polder
potentials \cite{scheelbuhmann08} and their resonant parts are

\begin{eqnarray}\label{eq:cpeone}
U_{e}(\mathbf{r}_{A})
= \frac{\hbar\mu_{0}}{2\pi}\int^{\infty}_{0}\mathrm{d}\xi\
\xi^2\mathrm{tr}[\bm{\alpha}(i\xi)\sprod\bm{G}^{(1)}(\mathbf{r}_{A},
\mathbf{r}_{A}, i\xi)] \nonumber \\-
\mu_{0}\sum_{k}\Theta(\omega_{nk})\omega^{2}_{nk}\mathbf{d}_{nk}
\sprod \mathrm{Re}[\bm{G}^{(1)}(\mathbf{r}_{A}, \mathbf{r}_{A},
\omega_{nk})] \sprod \mathbf{d}_{kn}
\end{eqnarray}and 
\begin{eqnarray}\label{eq:cpmone}
U_{m}(\mathbf{r}_{A})
= \frac{\hbar\mu_{0}}{2\pi}\int^{\infty}_{0}\mathrm{d}\xi\
\mathrm{tr}[\bm{\beta}(i\xi)\sprod\mathbf{\nabla}\vprod
\bm{G}^{(1)}(\mathbf{r}_{A},\mathbf{r}_{A},
i\xi)\vprod\overleftarrow{\mathbf{\nabla}}'] \nonumber \\+ \mu_{0}
\sum_{k} \Theta(\omega_{nk}) \mathbf{m}_{nk} \sprod
\mathbf{\nabla}\vprod \mathrm{Re}[\bm{G}^{(1)}(\mathbf{r}_{A},
\mathbf{r}_{A}, \omega_{nk})] \vprod \mathbf{\nabla}' \sprod
\mathbf{m}_{kn}
\end{eqnarray} where
\begin{equation}\label{eq:alpha}
\bm{\alpha}(i\xi)
=
\frac{1}{\hbar}\sum_{k}\left(\frac{\mathbf{d}_{kn}\tprod\mathbf{d}_{nk
}}{\omega_{kn}\!+\!i\xi} +
\frac{\mathbf{d}_{nk}\tprod\mathbf{d}_{kn}}{\omega_{kn}\!-\!i\xi}
\right)
\end{equation}
and
\begin{equation}\label{eq:beta}
\bm{\beta}(i\xi)
= \frac{1}{\hbar}\sum_{k}\left(
\frac{\mathbf{m}_{kn}\tprod\mathbf{m}_{nk}}{\omega_{kn}+i\xi}\!+\!
\frac{\mathbf{m}_{nk}\tprod\mathbf{m}_{kn}}{\omega_{kn}\!-\!i\xi}
\right)
\end{equation} 
are the usual electric and magnetic susceptibilities.


\section{Chiral Particle near a Chiral Halfspace}
\begin{figure}[t!]
\centering
\includegraphics{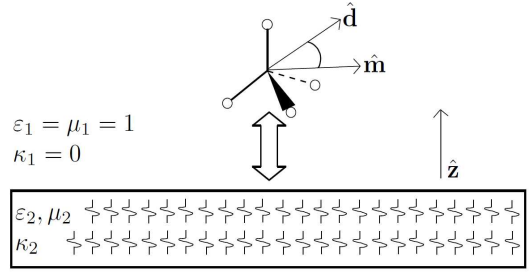}
\caption{Chiral molecule in free space near a chiral medium. The
chiral molecule will have a different group or base (denoted by
$\circ$) at the end of each bond, breaking the mirror symmetry.
The boundary between the
halfspaces is the ($x,y$) plane at $z = 0$. The isotropic chiral
medium fills the space with $z < 0$ (labelled region 2) and $z > 0$ is
the free-space (region 1).}
\label{figchiralparticle}
\end{figure}

First we consider a chiral particle in free space near halfspace
containing an isotropic chiral medium (figure
\ref{figchiralparticle}).
To compute the Casimir-Polder potential as given in (\ref{eqchiralcp})
the scattering part of the Green's function for a source in free space
near an isotropic chiral medium is required. The wave vectors
travelling in the achiral region 1 are the standard free-space
wave vectors
\begin{equation}  
k_{1}^{2} = \left(\frac{\omega}{c}\right)^{2},\quad k_{1z}^{2}
= k_{1}^{2} - k_{1q}^{2} 
\end{equation}%
where $k_{q}^{2} = k_{x}^{2} + k_{y}^{2}$. Within the chiral medium the
wavevectors are
\begin{equation}
(k^{R}_{2})^2 = \left(\frac{\omega}{c}\right)^{2}\!(\kappa_{2}
+ \sqrt{\varepsilon_{2}\mu_{2}})^{2}, 
\end{equation}\begin{equation}
(k^{L}_{2})^2 = \left(\frac{\omega}{c}\right)^{2}\!(-\kappa_{2}
+ \sqrt{\varepsilon_{2}\mu_{2}})^{2},
\end{equation}
\begin{equation}
(k^{P}_{2z})^2 = (k^{P}_{2})^2 - (k_{2q})^{2},
\end{equation}
where $P = R, L$ and refers to the right and left circular
polarisation of the wave, respectively.

The scattering part of the dyadic Green's function for a reflection
from the achiral/chiral interface can be found in
\cite{alihabashykong92} as
\begin{eqnarray}\label{eqgeneralgreen}
\bm{G}(\mathbf{r}, \mathbf{r}',\omega)\! =\! \frac{i}{8\pi^2} 
\frac{c}{\omega} \int
\mathrm{d}\mathbf{k_{1q}}\frac{k_{1}}{k_{1z}}e^{i(\mathbf{k_{1q}}\,
\sprod\,(\mathbf{r_{q}} - \mathbf{r_{q}}') + k_{1z}(z+z'))}
\biggl[\mathbf{e}_{s} \mathbf{e}_{s}R^{(s,s)} \nonumber \\ +
\mathbf{e}_{s} \mathbf{e}_{p}(-k_{1z})R^{(s,p)}+
\mathbf{e}_{p}(k_{1z})\mathbf{e}_{s}R^{(p,s)} +
\mathbf{e}_{p}(k_{1z}) \mathbf{e}_{p}(-k_{1z})R^{(p,p)}\biggr]
\end{eqnarray}
where 
\begin{equation}
\mathbf{e}_{s} =  \frac{1}{k_{q}}(k_y\mathbf{x} -
k_x\mathbf{y}) \end{equation}and\begin{equation}
\mathbf{e}_{p}(\pm k_z) = \frac{1}{k}(\mp k_z \mathbf{k}_{q}
+ k_q\mathbf{z})
\end{equation}
are the polarisation unit vectors for $s$ and $p$ polarised waves. The 
reflection coefficients $R^{(s,s)}, R^{(s,p)}, R^{(p,s)}$ and $R^{(p,p)}$ 
are typically dependent on the wave vectors given
above and the medium response functions.

In this geometry the chiral component for the Casimir-Polder potential is
\begin{eqnarray}\label{eqcpchiral}
U_{c}(z_{A})  &&= \frac{\hbar \mu_{0}}{4\pi^{2}c} \int^{\infty}_{0}
\mathrm{d}\xi\ \xi^{2} \Gamma(i\xi) \int^{\infty}_{\frac{\xi}{c}}
\mathrm{d} \tilde{k}_{1z} e^{-2\tilde{k}_{1z} z_{A}}
\nonumber \\ &&\vprod\bigg[\bigg(\frac{2\tilde{k}^{2}_{1z}c^{2}}{\xi^{2}}\!-\!1\bigg)R^{(s,
p)}(i\xi) + R^{(p,s)}(i\xi)\bigg] \nonumber \\ &&-
\frac{\mu_{0}}{12\pi^{2}c} \sum_{k} \Theta (\omega_{nk}) \omega^{2}_{nk}
R_{nk} \mathrm{Re}\int\mathrm{d}\mathbf{k}_{1q}  \frac{e^{i2k_{1z}
z_{A}}}{k_{1z}} \nonumber \\
&&\vprod \bigg[\bigg(\frac{2k^{2}_{1z}c^{2}}{\omega_{nk}^{2}}\!-\!1\bigg)R^{(s,
p)}(\omega_{nk}) + R^{(p,s)}(\omega_{nk})\bigg],
\end{eqnarray}
the details of this calculation can be found in \ref{appgreen}.

\subsection{Perfect Chiral Mirror}
\label{secperfect}
As a purely theoretical construct we consider the potential between an
isotropic molecule and an idealised medium that we have dubbed a
`perfect chiral mirror', whose reflection coefficients are $R^{(s,p)},
R^{(p,s)} = \pm 1$ and therefore $R^{(s,s)}, R^{(p,p)} = 0$. This
represents reflections where perpendicularly polarised waves are
completely reflected into parallel polarised waves and vice
versa. For a chiral medium that rotates the polarisation clockwise
(with regard to the incoming wave - labelled `right-handed') the
reflection coefficients are $R^{(s,p)} = 1$ and $R^{(p,s)} = - 1$ and
for an anticlockwise polarisation rotation (labelled `left-handed')
they are $R^{(s,p)} = - 1$ and $R^{(p,s)} = 1$.
When the molecule is initially in its ground state, applying these
reflection coefficients to (\ref{eqcpchiral}) results in
\begin{equation}\label{eqperfectchiralcp}
U(z_{A})\! =\! \pm  \frac{\hbar Z_0}{8\pi^{2}z_{A}^3}\!
\int^{\infty}_{0}\!\mathrm{d}\xi\Gamma(i\xi) e^{-2\frac{\xi}{c}z_A}\!
\biggl( \frac{2\xi z_{A}}{c}+ 1\biggl)
\end{equation}
where the `$+$' (upper sign) refers to the right-handed medium and
`$-$' (lower sign) refers to the left-handed medium. To examine how
the spatial separation between the chiral particle and the chiral
halfspace affects the chiral potential, we take the far-distance
(retarded) and close-distance (non-retarded) limits of
(\ref{eqperfectchiralcp}).

In the retarded limit where $z_{A}\omega_{\mathrm{min}}/c \gg 1$
($\omega_{\mathrm{min}}$ is the minimum relevant particle transition
frequency), the cross polarisability, $\Gamma(i\xi)$, approaches the
static limit and can be approximated by $\Gamma(i\xi)\approx
\Gamma'(0)\xi$. The potential becomes
\begin{equation}
U(z_{A}) =
\mp
\frac{Z_{0}c^{2}}{16\pi^{2}z_{A}^{5}}
\sum_{k}\frac{R_{0k}}{(\omega_{k0})^{2}}.
\end{equation}
In the non-retarded limit where $z_{A}\omega_{\mathrm{max}}/c \ll 1$
($\omega_{\mathrm{max}}$ is the maximum relevant particle transition
frequency) the potential becomes
\begin{equation}
U(z_{A}) = \pm \frac{Z_{0}}{12\pi^{2}z_{A}^{3}} \sum_{k} R_{0k}
\ln\left(\frac{\omega_{k0}z_{A}}{c} \right)
\end{equation}
where use has been made of the relationship 
\begin{equation*}
\int^{b}_{a}\mathrm{d}x \frac{x}{a^2 + x^2} \approx -\ln a,
\mathrm{for}\ a \ll 1.
\end{equation*}
The results show a spatial scaling behaviour in the retarded and
non-retarded limit that is different to the cases seen for dielectric
or magnetic media in this geometry. An interesting consequence of such
a medium is that in this simple geometry the electric and magnetic
components of the Casimir-Polder potential would vanish. It should be
noted that the potential can be attractive or repulsive depending on
the medium and particle in question, in contrast to the purely
attractive potentials that usually arise in this geometry. The chiral
identity of both objects determines the character of the potential.
The lack of electric and magnetic components of the Casimir--Polder
potential and the existence of repulsive forces means that perfect
chiral mirrors used in a cavity or Fabry-P\'erot geometry could be
used to separate enantiomers. We believe that the results represent
a theoretical upper bound for the chiral Casimir-Polder potential,
however, they are physically unrealisable. This is because it would
require a medium that completely rotates the polarisation of the
incident waves and perfectly reflects the waves. Typically, the
chirality of a medium will be restricted to $\kappa^{2} <
\varepsilon\mu$, which would exclude a perfect chiral mirror.

\subsection{Isotropic Chiral Medium}
\label{secmeta}
For all realistic media, the reflection coeficients are not unity as
there will always be transmission and losses from the fields as they
are reflected. The reflection coefficients for an isotropic medium are
given in \cite{alihabashykong92}, after taking the non-retarded limit
they are
\begin{equation}
R^{(s,p)} = - R^{(p,s)} =
\frac{2i\kappa_{2}}{\varepsilon_{2}\mu_{2}\! -\! \kappa_{2}^{2}\!
+\! \varepsilon_{2}\! +\! \mu_{2}\! +\! 1},
\end{equation}\begin{equation}
R^{(s,s)} =  \frac{\varepsilon_{2}\mu_{2}\! -\! \kappa_{2}^{2}\! -\!
\varepsilon_{2}\! +\! \mu_{2}\! -\! 1}{\varepsilon_{2}\mu_{2}\! -\!
\kappa_{2}^{2}\! +\! \varepsilon_{2}\! +\! \mu_{2}\! +\! 1}
\end{equation}and\begin{equation}
R^{(p,p)} =  \frac{\varepsilon_{2}\mu_{2}\! -\! \kappa_{2}^{2}\! +\!
\varepsilon_{2}\! -\! \mu_{2}\! -\! 1}{\varepsilon_{2}\mu_{2}\! -\!
\kappa_{2}^{2}\! +\! \varepsilon_{2}\! +\! \mu_{2}\! +\! 1}.
\end{equation}
By setting $\kappa_2 = 0$ the cross reflection coefficients disappear
and $R^{(s,s)}, R^{(p,p)}$ revert to the standard Fresnel reflection
coefficients in the non-retarded limit.

We can now obtain the chiral-corrected electric component and the
chiral component of the Casimir-Polder potential in the non-retarded
limit. The off-resonant terms read
\begin{eqnarray}\label{eqecphm}
U_{eO}(z_{A}) =  -\frac{\hbar}{16\pi^2\varepsilon_{0}z_{A}^3}
&&\int^{\infty}_{0} \mathrm{d}\xi\ \alpha (i\xi) \nonumber \\ &&\vprod
\biggl[\frac{\varepsilon_{2}\mu_{2}\! -\! \kappa_{2}^{2}\! +\!
\varepsilon_{2}\! -\!
\mu_{2}\!-\!1}{\varepsilon_{2}\mu_{2}\!-\!\kappa_{2}^{2}\!
+\!\varepsilon_{2}\! +\! \mu_{2}\! +\! 1}\biggl],
\end{eqnarray}
\begin{equation}\label{eqccphm}
U_{cO}(z_{A}) = \frac{\hbar Z_{0}}{4\pi^2z_{A}^3}
\int^{\infty}_{0}\mathrm{d}\xi\ \Gamma(i\xi)
\biggl[\frac{i\kappa_{2}}{\varepsilon_{2}\mu_{2}\! -\!
\kappa_{2}^{2}\! +\! \varepsilon_{2}\! +\! \mu_{2}\! +\!
1}\biggl],
\end{equation}
where all response functions are dependent on complex frequency, i.e.,
$\varepsilon_{2} = \varepsilon_{2}(i\xi), \mu_{2} = \mu_{2}(i\xi)$ and
$\kappa_{2} = \kappa_{2}(i\xi)$.
The resonant terms are
\begin{eqnarray}\label{eqecpres}
U_{eR}(z_{A}) =  -\frac{1}{24\pi\varepsilon_{0}z_{A}^3}&& \sum_{k}
\Theta(\omega_{nk})\vert d_{nk}\vert^{2} \nonumber \\ &&\vprod
\mathrm{Re}\biggl[\frac{\varepsilon_{2}\mu_{2}\! -\!
\kappa_{2}^{2}\! +\! \varepsilon_{2}\! -\! \mu_{2}\! -\!
1}{\varepsilon_{2}\mu_{2}\! -\! \kappa_{2}^{2}\! +\!
\varepsilon_{2}\! +\! \mu_{2}\! +\! 1}\biggl],
\end{eqnarray}
\begin{eqnarray}\label{eqccpres}
U_{cR}(z_{A}) =  \frac{Z_{0}}{6\pi z_{A}^3} &&\sum_{k}\Theta
(\omega_{nk}) R_{nk} \nonumber \\
&&\vprod\mathrm{Im}\biggl[\frac{i\kappa_{2}}{\varepsilon_{2}\mu_{2}\!
-\! \kappa_{2}^{2}\! +\! \varepsilon_{2}\! +\! \mu_{2}\! +\!
1}\biggl],
\end{eqnarray}
where the response functions in (\ref{eqecpres}) and (\ref{eqccpres})
are taken at the transition frequency, $\omega_{nk}$ i.e.,
$\varepsilon_{2} = \varepsilon_{2}(\omega_{nk}), \mu_{2} =
\mu_{2}(\omega_{nk})$ and $\kappa_{2} = \kappa_{2}(\omega_{nk})$.
As a consistency check, setting $\mu_{2} = 1$ and letting $\kappa_{2}
\rightarrow 0$ in the off-resonant contribution the results for the
electric component of the Casimir-Polder potential between a particle
and a dielectric in \cite{scheelbuhmann08} are recovered.


\section{Chiral Cavity}
To examine whether the Casimir-Polder force can be used to
distinguish between enantiomers, we consider a chiral molecule in free
space between two chiral media which differ only in their handedness.
To illustrate this we have selected a chiral metamaterial whose
parameters have been published in \cite{rillopplett09} as the chiral
medium due to the strong chirality that can be obtained from
metamaterials. To characterise the medium a single-resonance
Drude-Lorentz model is used for $\varepsilon (\omega)$ and $\mu
(\omega)$ and the Condon model \cite{condon37, tsitsas2011} is used
for $\kappa (\omega)$. These are
\begin{equation}\label{eqepsilon}
\varepsilon(\omega)  = 1 - \frac{\omega_{p}^{2}}{\omega^{2} -
\omega_{E}^{2} + i\gamma_{E}\omega},
\end{equation}
\begin{equation}\label{eqmu}
\mu (\omega)  = 1 - \frac{\omega_{m}^{2}}{\omega^{2} - \omega_{B}^{2}
+ i\gamma_{B}\omega}
\end{equation}
and
\begin{equation}\label{rho}
\kappa (\omega)  = \frac{a\omega}{\omega^{2} - \omega_{C}^{2} +
i\gamma_{C}\omega}.
\end{equation}
The constants $\omega_{p}, \omega_{m}$ and $a$ represent the
oscillator strengths for the dipole transitions responsible for
$\varepsilon$, $\mu$ and the rotatory strength associated with
$\kappa$, respectively. The remaining constants are the resonant
frequencies and damping factors for the permittivity ($\omega_{E}$,
$\gamma_{E}$), permeability ($\omega_{B}$, $\gamma_{B}$) and the
chirality ($\omega_{C}$, $\gamma_{C}$) of the medium.

For the chiral molecule, dimethyl disulphide
$(\mathrm{CH}_{3})_{2}\mathrm{S}_{2}$ has been chosen. The dipole
and rotatory strengths for each transition have been numerically
calculated for various orientations \cite{rauk84}. As an example, we
have chosen the first transition when the orientation between the two
$\mathrm{CH}_{3} - \mathrm{S} - \mathrm{S}$ planes is $90^{\circ}$.
The transition frequency is $\omega_{nk} = 9.17 \mathrm{x} 10^{15} s^{-1}$, 
dipole strength $\vert \mathbf{d}_{nk} \vert^{2} = 8.264
\mathrm{x} 10^{-60}$ (C m)$^{2}$ and the rotatory strength is $R_{nk}
= 3.328 \mathrm{x} 10^{-64}$ C$^{2}$ m$^{3}$ s$^{-1}$.

\subsection{Ground-State Force}
\label{secground}
In the non-retarded limit, multiple reflections between the halfspaces
are not considered and the forces on the molecule for each halfspace
are simply added. From the Casimir-Polder potentials for the electric
(\ref{eqecphm}) and chiral (\ref{eqccphm}) components we can obtain
the forces ($\mathbf{F} = -\mathbf{\nabla}U$) acting on the molecule.
The two halfspaces are identical except for their chirality,
characterised by $a$, which is negative for one of the halfspaces and
positive for the other. We assume that the molecule has only two
energy levels, i.e. only one transition is considered.

\begin{figure}[t!]
\centering
\includegraphics{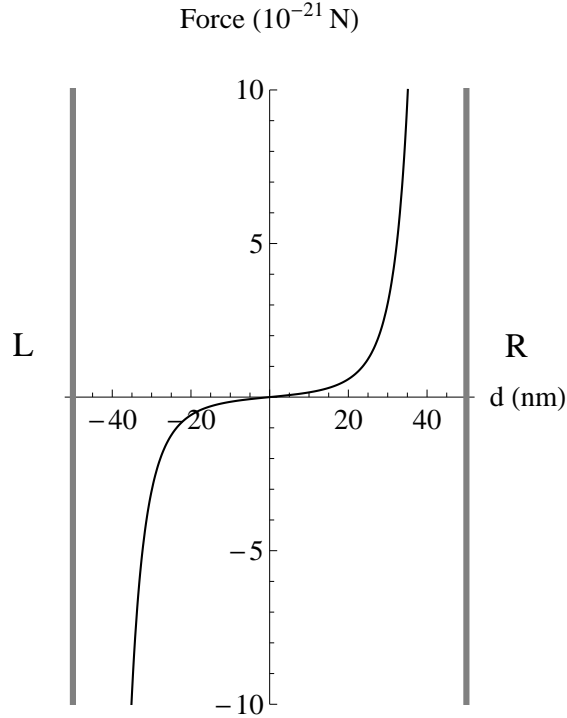}
\caption{The electric component of the Casimir-Polder force. The
molecule is attracted to both halfspaces with the stronger attraction
coming from the closest halfspace. The vertical lines denote the
boundaries of the metamaterial, the gap between is freespace where the
molecule is located. The parameters obtained from \cite{rillopplett09}
are $\omega_{p} = 5.47 \mathrm{x} 10^{14} s^{-1}$, $\omega_{m} = 3.06
\mathrm{x} 10^{14} s^{-1}$, $a = -3.61 \mathrm{x} 10^{14} s^{-1}$, $\omega_{E} =
\omega_{B} = \omega_{C} = 4.96 \mathrm{x} 10^{14} s^{-1}$, $\gamma_{E} =
\gamma_{B} = 2.51 \mathrm{x} 10^{13} s^{-1}$ and $\gamma_{C} = -2.58
\mathrm{x} 10^{13} s^{-1}$.}
\label{figeleccpforce}
\end{figure}

\begin{figure}[t!]
\centering
\includegraphics{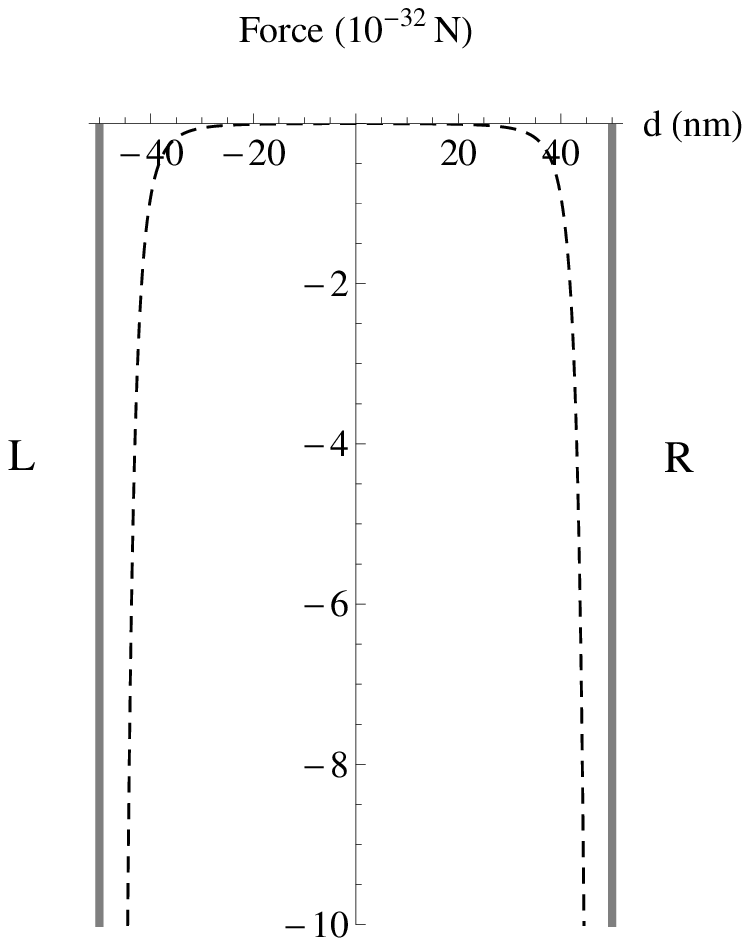}
\caption{The chiral component of the Casimir-Polder force. The
molecule is attracted to the left hand side halfspace and there is a
repulsive force between the molecule and the right hand side
halfspace.}
\label{figchircpforce}
\end{figure}

By considering a separation between the halfspaces of $100
\mathrm{nm}$ we obtain the results shown in figure
\ref{figeleccpforce} and figure \ref{figchircpforce}.

The electric component of the Casimir-Polder force (figure
\ref{figeleccpforce}) is always attractive towards both halfspaces. In
this geometry these forces are equal and opposite, so when computing
the electric component of the total force (in the non-retarded limit)
on the molecule, the halfspace closest to the molecule will provide
the dominant contribution to this total force. By implication, when
the molecule is in the centre (i.e. equal distance from both
halfspaces), the electric components of the Casimir-Polder force
cancel and the net contribution to the total Casimir-Polder force is
zero. It should be noted that the difference in the chirality of the
halfspaces does not effect the electric force from either halfspace
and hence the combined electric component of the total force.

By looking at the chiral component of the Casimir-Polder force (figure
\ref{figchircpforce}) one obtains an attractive force between the
chiral molecule and one of the halfspaces (to the left hand side) and
a repulsive force between the chiral molecule and the other halfspace
(to the right hand side). Furthermore, the total chiral component of
the force does not disappear at the midpoint between the halfspaces.

Figure \ref{figeleccpforce} and figure \ref{figchircpforce} show that
the electric component of the Casimir-Polder force is many orders of
magnitude larger than the chiral component and will dominate
interactions. The exception to this is the central region between the
halfspaces, where the overall electric component is reduced
sufficiently to allow the chiral component to become the dominant
force. However, the width of this central region is smaller than the
molecule. This means that for a particle initially in the ground state
the Casimir-Polder force, in the current geometry, would not be able
to distinguish between enantiomers and subsequently separate them.

The difference in orders of magnitude of the force components can be
traced to separate origins. With regard to the chiral molecule the
optical rotatory strength is orders of magnitude smaller than the
electric dipole transition matrix element, 
$R_{nk}/c\lesssim 10^{-11} |\mathbf{d}_{nk}|^2$. This can be understood by
the fact that the magnetic dipole moment appears at a higher-order of
the multipole expansion than the electric dipole moment. Looking at
the chiral medium, the chirality is slightly smaller than the
permittivity, but this difference is enhanced by the structure of the
reflection coefficients.

\subsection{Excited-State Force}
\label{secexcited}
In order to overcome the dominance of the electric dipole force, it is
helpful to consider other molecular initial states. If the molecule is
initially in an excited state the resonant contribution to the force
needs to be considered. In this scenario, the off-resonant part of the
electric component of the Casimir-Polder force is repulsive whereas
the direction of the chiral off-resonant component is dependent on the
chiral identity of the molecule and material in question. A
metamaterial can be constructed such that for a given transition
frequency the resonant contribution to the electric component of the
Casimir-Polder force will counteract the off-resonant contribution. By
creating a cavity with two such metamaterials, identical except for
their chirality, a chiral molecule initially in an excited state can
be separated from its enantiomer. This is because the suppression of
the electric component of the Casimir-Polder force allows the chiral
component of the force to attract an enantiomer to the corresponding
halfspace while it is being repelled from the opposite halfspace
without the electric component dominating.

As a proof of principle the metamaterials parameters are
chosen as suitable multiples of the plasma frequency and the values for 
the optical rotatory strength
and dipole strength are as given above and the metamaterials are
separated by $100$ nm of free space.

\begin{figure}[t!]
\centering
\includegraphics{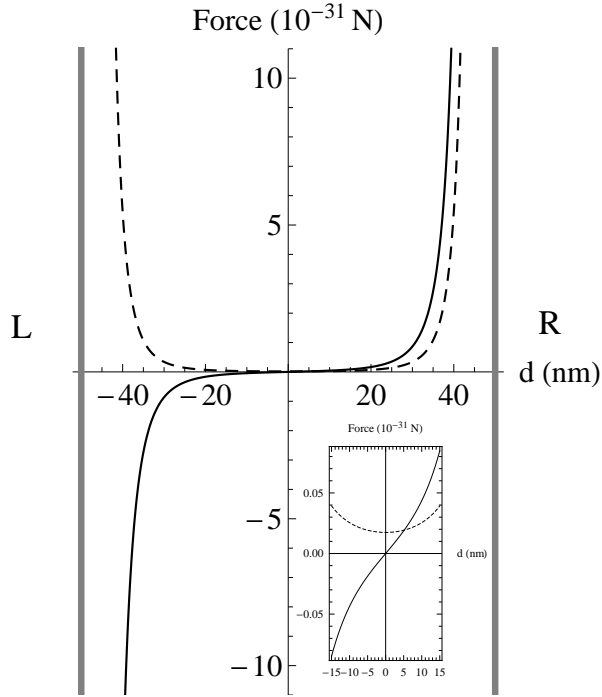}
\caption{The total electric component of the Casimir-Polder force
(black line) and the total chiral component (dashed line). The
components of the force are of equal magnitude, with the chiral
contribution larger in the centre and the electric contribution larger
towards the halfspaces. The parameters are chosen as $\omega_{E} =
\omega_{B} = \omega_{C} = \frac{\omega_{p}}{2}$, $\gamma_{E} =
\gamma_{B} = - \gamma_{C} = \frac{\omega_{p}}{10^{3}}$, $a = -
\frac{\omega_{p}}{3}$ and $\omega_{m} = \frac{\omega_{p}}{5}$. The
inset in the bottom right shows a magnification of the centre region
between $-15$ nm and $15$ nm.}
\label{figresonances}
\end{figure}
The results are shown in figure \ref{figresonances}. As can be seen,
for a particular transition frequency a suppression of the electric
component of the Casimir-Polder force sufficient to allow enantiomer
separation can be obtained. The chiral component is the dominant
contribution in the central region ($\approx 10 \mathrm{nm}$) of the
cavity (see the inset to figure \ref{figresonances}), meaning that the
direction of the force acting on a chiral molecule in this region will
be dependent on its chirality. Therefore enantiomers that pass at low
speeds through the centre of the cavity will be attracted or repelled
in opposite directions and will be separated based on their chirality.
As the total electric component of the Casimir-Polder force is
attractive, the separated enantiomers will continue to be drawn
towards opposite halfspaces even when not in the central region.

It is important to note that, although the chirality of the
metamaterials has not changed, the chiral component of the
Casimir-Polder force is now acting in the opposite direction. This is
due to the resonant part of the chiral force, which is larger than the
off-resonant part and acts against it.


\section{Conclusion}
\label{secconclusion}
The chiral component of the Casimir-Polder potential has been derived
within the framework of macroscopic QED. The results show that the
medium and particle in question must both be chiral, otherwise this
potential does not exist. Furthermore, this potential is sensitive to
the chirality of the objects in question and can be attractive or
repulsive.

By initially considering a perfect chiral mirror, it was found that in
the retarded limit the chiral component of the Casimir-Polder
potential scales as $1/(z_{A})^{5}$ with the molecule-surface distance
whereas for the non-retarded limit the spatial scaling is
$1/(z_{A})^{3}\ln (z_{A})$. As already alluded to, it is unlikely that this could
be realised in a real material, due to the requirement for perfect
reflection {\it and} complete rotation of the incident wave
polarisation. When the chiral medium does not exhibit perfect
reflectance the chiral potential is slightly diminished and in the
non-retarded limit the spatial scaling was found to be
$1/(z_{A})^{3}$.

In the geometry where a chiral molecule, initially in the ground
state, is located between two half spaces of opposite chirality it was
found that the chiral component of the Casimir-Polder force is
attractive towards one half space and repulsive from the other.
However, the electric component of the Casimir-Polder force between
the molecule and the half spaces dominates over the chiral component.
Therefore, in this geometry, it would not be possible to distinguish
between enantiomers because the indiscriminate attractive electric 
force dominates over any chiral effects.

If the molecule is initially in an excited state it was found that the
material properties can be tuned such that the resonant contribution
to the electric component of the Casimir-Polder force almost completely
suppresses the off-resonant contribution. Enantiomers in the centre
between the halfspaces would then experience an overall force whose
direction is dependent on the chiral identity of the molecule. This
chiral force will draw the molecule towards a particular halfspace
with its enantiomer attracted to the other side of the cavity. This
distinction between enantiomers will allow them to be separated by the
Casimir-Polder force.

\ack
The authors would like to thank K. Hornberger and J. Trost for useful
discussions. The research was funded by the UK Engineering and
Physical Sciences Research Council (EPSRC).

\appendix
\section{}
\section*{Integral Relation}
\label{appintrel}

We begin by evaluating the left hand side of the integral
equation~(\ref{intG}).
By taking the product of (\ref{eqgegm}) with its hermitian conjugate
we arrive at the term
\begin{eqnarray}
\bm{G}_{e}\conv\bm{G}^{\dag}_{e}
+ \bm{G}_{m} \conv\bm{G}^{\dag}_{m}
= \nonumber \\
(\mu_{0}\omega)^{2}\frac{\hbar}{\pi} \bm{G}
\conv
\left(\begin{array}{c}i\omega \\
\vprod\overleftarrow{\mathbf{\nabla}}'\end{array}\right) \sprod
\mathcal{R} \conv \mathcal{R}^{\dag}
\sprod \left(\begin{array}{c}i\omega \\
\vprod\overleftarrow{\mathbf{\nabla}}'\end{array}\right)^{\dag}\conv
\bm{G}^{\dag}.
\end{eqnarray}
Substituting in (\ref{eqrrdagmatrix}) leads to
\begin{eqnarray}\label{eqintrellong}
\sum_{\lambda = e,m} \bm{G}_{\lambda}
\conv \bm{G}^{\dag}_{\lambda} = 
\frac{\hbar\mu_{0}\omega^{2}}{\pi}\Bigl[\bm{G}
\conv\Bigl[\frac{\omega^{2}}{c^{2}}
\mathrm{Im}[\bm{\varepsilon} \!-\!\bm{\kappa}^{\mathrm{T}}
\conv\bm{\mu}^{-1} \conv\bm{\kappa}] \nonumber
\\ + \frac{\omega}{c}\bigl(\vprod \overleftarrow{\mathbf{\nabla}}'
\!\sprod\! \mathrm{Im}[\bm{\mu}^{-1} \conv\bm{\kappa} ]\! -\!
\mathrm{Im}[\bm{\kappa}^{\mathrm{T}} \conv\bm{\mu}^{-1}
] \!\sprod\! \mathbf{\nabla} \vprod\bigl) \nonumber \\ +
\vprod \overleftarrow{\mathbf{\nabla}}' \!\sprod\!
\mathrm{Im}[\bm{\mu}^{-1}] \!\sprod\! \mathbf{\nabla}
\vprod\Bigl]\, \conv\, \bm{G}^{\dag} \Bigl].
\end{eqnarray}
By comparing (\ref{eqintrellong}) with the noise polarisation and
magnetisation commutators (\ref{eqjnjn}), (\ref{eqjnmn}),
(\ref{eqmnjn}) and (\ref{eqmnmn}) it can be seen that this reduces to
\begin{equation}
\sum_{\lambda = e,m} \bm{G}_{\lambda}(\omega)\conv
\bm{G}^{\dag}_{\lambda} (\omega) =
\frac{\hbar\mu_{0}\omega^{2}}{\pi}\Bigl[\bm{G}(\omega) \conv
\Bigl[\hat{\mathbf{j}}_\mathrm{N}(\omega),
\hat{\mathbf{j}}_\mathrm{N}^\dagger (\omega)\Bigr] \conv
\bm{G}^{\dag}(\omega) \Bigl].
\end{equation}
It is known that \cite{njp_2012}
\begin{equation}
\Bigl[\hat{\mathbf{j}}_\mathrm{N}(\mathbf{r},\omega),
\hat{\mathbf{j}}_\mathrm{N}^{\dag}(\mathbf{r}',\omega')\Bigr]
=\frac{\hbar\omega}{\pi}\,\mathrm{Re}
[\bm{Q}(\mathbf{r},\mathbf{r}',\omega)] \delta(\omega-\omega')
\end{equation}
where $\bm{Q}$ is a complex conductivity tensor. This leads to the
general form of the integral relation \cite{njp_2012}
\begin{eqnarray}
\left(\frac{\hbar\mu_{0}\omega^{2}}{\pi}\right)\mu_{0}\omega
[\bm{G}(\omega)\conv
\mathrm{Re}\bm{Q}(\omega)\conv\bm{G}^{\dagger}(\omega)]
(\mathbf{r},\mathbf{r}') \nonumber \\
=
\frac{\hbar\mu_{0}\omega^{2}}{\pi}\mathrm{Im}\bm{G}
(\mathbf{r},\mathbf{r}',\omega).
\end{eqnarray}
Therefore, 
\begin{equation}
\sum_{\lambda}[\bm{G}_{\lambda}(\omega)
\conv\bm{G}^{\dag}_{\lambda}(\omega)](\mathbf{r}, \mathbf{r}') =
\frac{\hbar\mu_{0}\omega^{2}}{\pi}\mathrm{Im}
\bm{G}(\mathbf{r},\mathbf{r}',\omega),
\end{equation}
which in coordinate space is
\begin{equation}
\sum_{\lambda}\int\!\mathrm{d}^{3}\mathbf{s}
\bm{G}_{\lambda}(\mathbf{r}, \mathbf{s},
\omega)\sprod\bm{G}^{\dag}_{\lambda}(\mathbf{r}', \mathbf{s}, \omega)
\!=\!\frac{\hbar\mu_{0}\omega^{2}}{\pi} \mathrm{Im}\bm{G}(\mathbf{r},
\mathbf{r}', \omega),
\end{equation}
as required.

\section{}
\section*{Dyadic Green's Function Calculation}
\label{appgreen}

To obtain the chiral component of the Casimir-Polder force we require the
terms for $\mathrm{tr}[\mathbf{\nabla}\vprod \bm{G}(\mathbf{r}_{A},
\mathbf{r}_{A}, i\xi)]$ and $\mathrm{tr}[\mathbf{\nabla} \vprod
\mathrm{Re}[\bm{G} (\mathbf{r}_{A}, \mathbf{r}_{A}, \omega_{nk})]]$.
It is known from dyadic algebra that \cite{hcchen_emwaves} 
\begin{equation}\label{eqdyad}
\mathbf{u}\vprod \mathbf{a}\mathbf{b} = (\mathbf{u}\vprod
\mathbf{a})\mathbf{b}
\end{equation} where $\mathbf{a}\mathbf{b}$ is a dyadic product and
subsequently $(\mathbf{u}\vprod \mathbf{a})\mathbf{b}$ is also a
dyadic product. 
Applying (\ref{eqdyad}) to (\ref{eqgeneralgreen}) gives the effective
results 
\begin{equation}
\mathbf{\nabla}\vprod\mathbf{e}_{s} = - ik\mathbf{e}_{p}(k_{z}),
\end{equation}\begin{equation}
\mathbf{\nabla}\vprod\mathbf{e}_{p}(k_z) = ik\mathbf{e}_{s},
\end{equation}
where $\mathbf{\nabla} \rightarrow (ik_{x}, ik_{y}, ik_{z})$. Taking 
the trace of the dyadic Green's function is equivalent to taking the
dot product between the dyads,
\begin{equation}
\mathbf{e}_{s}\sprod \mathbf{e}_{p}(-k_{z})
= \mathbf{e}_{p}(k_{z}) \sprod \mathbf{e}_{s} = 0,
\end{equation}\begin{equation} 
\mathbf{e}_{s}\sprod\mathbf{e}_{s} = 1,
\quad \mathbf{e}_{p}(k_{z}) \sprod \mathbf{e}_{p}(-k_{z}) =
1\!-\!\frac{2k^{2}_{z}}{k^{2}}.
\end{equation}
To calculate the off-resonant contribution the integration variable is
changed to
\begin{equation*}
\int\mathrm{d}\mathbf{k_{q}} \rightarrow
\int^{2\pi}_{0}\mathrm{d}\theta\int^{\infty}_{0}\mathrm{d}k_{q} k_{q}
\rightarrow - 2\pi \int^{\infty}_{\frac{\omega}{c}}\mathrm{d}k_{z}
k_{z},
\end{equation*} 
the wavevectors in complex frequency ($\omega \rightarrow i\xi$) become 
\begin{equation}
k_{1} = i\frac{\xi}{c} = i\tilde{k}_{1}, k_{1z} = i\tilde{k}_{1z} 
\end{equation}with
\begin{equation}\tilde{k}_{1z}
= \sqrt{\left(\frac{\xi}{c}\right)^2 + \left(k_{1q}\right)^2}
\end{equation}
in the achiral halfspace and
\begin{eqnarray}
&&k_{2}^{R} = i\left(\frac{\xi}{c}\right)\left( \kappa_{2}
+ \sqrt{\varepsilon_{2}\mu_{2}}\right)= i\tilde{k}_{2}^{R}, \nonumber \\
&&k_{2}^{L} = i\left(\frac{\xi}{c}\right)\left( -\kappa_{2}
+ \sqrt{\varepsilon_{2}\mu_{2}}\right)= i\tilde{k}_{2}^{L}, \nonumber \\
&&k^{P}_{2z} = i\tilde{k}^{P}_{2z}, \nonumber \\ &&\tilde{k}^{P}_{2z} =
\sqrt{\left(\tilde{k}^{P}_{2} \right) + \left( k_{2q} \right)^2}
\end{eqnarray} 
in the chiral halfspace. 

\end{document}